\documentclass{raa}

\usepackage{amsmath}
\usepackage{graphicx}
\usepackage{stmaryrd}
\usepackage{txfonts}

\usepackage{url}
\usepackage{subfigure}
\usepackage{longtable}
\usepackage{lscape}
\usepackage{natbib}

\begin{document}

   \title{Star formation associated with the infrared dust bubble N68
}

   \volnopage{Vol.0 (201x) No.0, 000--000}      
   \setcounter{page}{1}          

   \author{Chuan Peng Zhang
      \inst{1,2,3,}\thanks{zcp0507@gmail.com}
   \and Jun Jie Wang
      \inst{1,2}
   }

   \institute{National Astronomical Observatories, Chinese Academy of Sciences,
              100012 Beijing, PR China
         \and
             NAOC-TU Joint Center for Astrophysics, 850000 Lhasa, PR China
         \and
             University of Chinese Academy of Sciences, 100080 Beijing, PR China
             }

   \date{Received~~2009 month day; accepted~~2009~~month day}

\abstract{We investigated the environment of the infrared dust
bubble N68 and searched for evidence of triggered star formation in
its surroundings. We performed a multiwavelength study of the nebula
with data taken from several large-scale surveys: GLIMPSE, MIPSGAL,
IRAS, NVSS, GRS, and JCMT. We analyzed the spectral profile and the
distribution of the molecular gas ($^{13}$CO J = 1 - 0 and J = 3 -
2), and the dust in the environment of the N68. The
position-velocity diagram clearly shows that the N68 may be
expanding outward. We used two three-color images of the
mid-infrared emission to explore the physical environment, and one
color-color diagram to investigate the distribution of young stellar
objects (YSOs). We found that the 24 $\mu$m emission is surrounded
by the 8.0 $\mu$m emission. Morphologically, the 1.4 GHz continuum
correlates strongly with the 24 $\mu$m emission, and the $^{13}$CO J
= 1 - 0 and J = 3 - 2 emissions correlate well with the 8.0 $\mu$m
emission. We investigated two compact cores located at the shell of
the N68. The spectral intensity ratios of $^{13}$CO J = 3 - 2 to J =
1 - 0 range from 5 to 0.3. In addition, young star objects, masers,
IRAS, and UC HII regions distribute at the shell of bubble. The
active region may be triggered by the expanding of the bubble N68.
 }

   \authorrunning{C. P. Zhang \& J. J. Wang}            
   \titlerunning{Star formation associated with the infrared dust bubble N68}  

   \maketitle

%
%
\section{Introduction}    
\label{sect:intro}

There are many signatures of star formation, for example,
outflow/inflow, dark cloud and HII region, which can be used to
investigate the process of star formation. Also there are many
different kinds of interactions which can trigger the star
formation, such as cloud-cloud collision\citep{li2012}, supernova
explosion \citep{xujl2011,xu2011}, bubble expansion \citep{s51} and
so on. \citet{chur2006,chur2007} detected and cataloged about 600
mid-infrared dust (MIR) bubbles between longitudes -60$^{\circ}$ and
+60$^{\circ}$. The bubbles have bright 8.0 $\mu$m shells that
enclose bright 24 $\mu$m interiors. The infrared (IR) dust bubbles
may be produced by exciting O- and/or B-type stars, which is located
inside bubble. The ultraviolet (UV) radiation from exciting stars
may heat dust and ionize the gas to form an expanding bubble shell
\citep{wats2008}, which is known as ''collect-and-collapse''
process. This process can trigger a massive star formation near the
shell clumps. Therefore, the bubbles present an important
opportunity to study the interaction between the HII regions and
molecular clouds.

A few individual bubbles have been studied well, such as N49 and
S51. There are many models and observations to explain the dusty
wind-blown bubbles, such as bubble N49 of \citet{ever2010}.
Recently, we reported an expanding bubble S51 shown with a shell and
a front side, employing $^{13}$CO and C$^{18}$O J = 1 - 0 emission
lines \citep{s51}. \citet{beau2010} reported CO J = 3 - 2 maps of 43
Spitzer identified bubbles. \citet{wats2008} present an analysis of
wind-blown, parsec-sized, mid-infrared bubbles and associated star
formation.

To add the investigated bubble examples, we selected the IR dust
bubble [CPA2006] N68 (hereafter N68) from the catalog of
\citet{chur2006}. N68 is a complete or closed ring and centered on
Galactic $l$=35.654, $b$=-0.062 (or R.A.(J2000) =
18$^{h}$56$^{m}$25$^{s}$.70, DEC.(J2000) =
02$^{\circ}$26$'$01$''$.0). It has a distance of 10.6 kpc, which was
obtained by \citet{ande2009} based on comparing the velocity of the
ionized gas with the maximum velocity of HI absorption, and one
looking for HI absorption at the velocity of molecular emission. In
addition, its size is 34 pc $\times$ 17 pc and the eccentricity of
the ellipse is 0.72.

In this work, we mainly present a multiwavelength study of the
environment surrounding the IR dust bubble N68. We explore its
surrounding interstellar medium (ISM) and search for signatures of
star formation. The observations and data are described in
Sect.\ref{sect:data}; the results and discussions about the bubble
N68 environment are presented in Sect.\ref{sect:results};
Sect.\ref{sect:summary} summarizes the results.

\section{Data}
\label{sect:data}

\begin{table*}[hp]
\caption{YSOs around the bubble N68} \label{table:YSO} \centering
\scriptsize
\begin{tabular}{cccc}
\hline \hline
Source   &  Name   &   R.A.(J2000)  &   DEC.(J2000)        \\
 & & $^{h}$~~~$^{m}$~~~$^{s}$ & $^{\circ}$~~~$'$~~~$''$     \\
\hline
IRAS    & IRAS18538+0222  &    18 56 21.30 &  +02 26 37.00 \\
IRAS    & IRAS18538+0216  &    18 56 23.50 &  +02 20 38.00 \\
IRAS    & IRAS18540+0220  &    18 56 35.60 &  +02 24 54.00 \\
CHII    & G35.590-0.025   &    18 56 22.67 &  +02 21 14.58 \\
EGO     & G35.040-0.470   &    18 57 03.30 &  +02 21 50.00 \\
Maser   & G35.578-0.031   &    18 56 22.55 &  +02 20 28.10 \\
UCHII   & G35.578-0.031   &    18 56 22.64 &  +02 20 26.30 \\
\hline
\end{tabular}

\end{table*}

We analyzed IR and millimeter wavelength data extracted from several
large-scale surveys including GLIMPSE \citep{benj2003,chur2009},
MIPSGAL \citep{care2009}, IRAS \citep{neug1984}, NVSS
\citep{cond1998}, GRS\footnote{This publication makes use of
molecular line data from the Boston University-FCRAO Galactic Ring
Survey (GRS). The GRS is a joint project of Boston University and
Five College Radio Astronomy Observatory, funded by the National
Science Foundation under grants AST-9800334, AST-0098562,
AST-0100793, AST-0228993, \& AST-0507657.}, and JCMT\footnote{The
James Clerk Maxwell Telescope is operated by the Joint Astronomy
Centre on behalf of the Science and Technology Facilities Council of
the United Kingdom, the Netherlands Organisation for Scientific
Research, and the National Research Council of Canada.}.

GLIMPSE is a MIR survey of the inner Galaxy performed with the
Spitzer Space Telescope. We used the mosaicked images from the
GLIMPSE and the GLIMPSE Point-Source Catalog (GPSC) in the
Spitzer-IRAC (3.6, 4.5, 5.8, and 8.0 $\mu$m). IRAC has an angular
resolution of between $1.5''$ and $1.9''$ \citep{fazi2004,wern2004}.
MIPSGAL is a survey of the same region as the GLIMPSE, using the
MIPS instrument on Spitzer. The MIPSGAL resolution is 6$''$ at 24
$\mu$m. The 60 and 100 $\mu$m fluxes of the IR Astronomical
Satellite (IRAS) were employed to estimate the Lyman continuum
ionizing flux. The NRAO VLA Sky Survey (NVSS) is a 1.4 GHz continuum
survey covering the entire sky north of -40 deg declination
\citep{cond1998}.

\begin{figure}
\centering
\includegraphics[width=0.49\textwidth, angle=0]{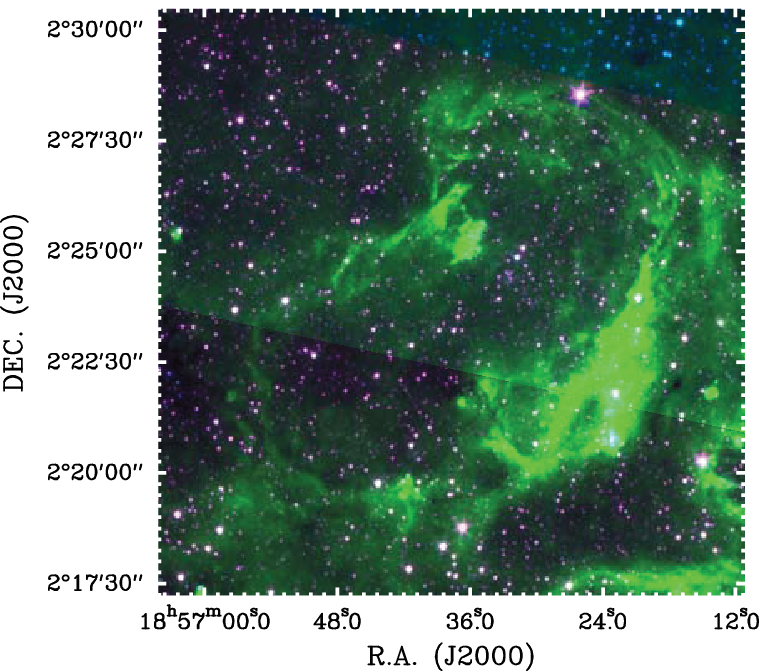}
\includegraphics[width=0.49\textwidth, angle=0]{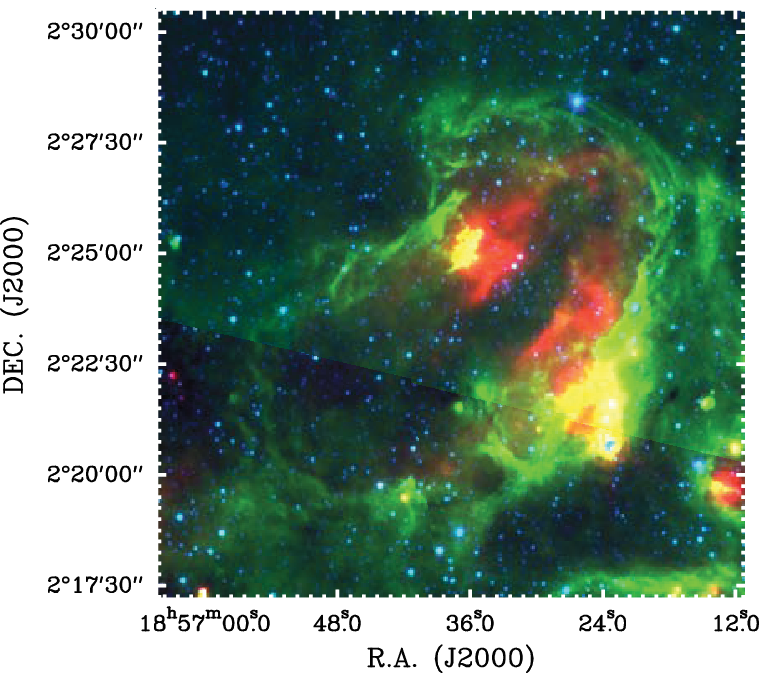}
\caption{Mid-IR emission of the IR dust bubble N68. $Left$:
Spitzer-IRAC three-color image (3.6 $\mu$m = red, 4.5 $\mu$m = blue,
and 8.0 $\mu$m = green); $Right$: Spitzer-IRAC and Spitzer-MIPSGAL
three-color image (4.5 $\mu$m = blue, 8.0 $\mu$m = green, and 24
$\mu$m = red).} \label{Fig:3_colors}
\end{figure}

The 14-m GRS has a full width at half maximum (FWHM) beam size of
$\sim46''$ at $\sim$110 GHz for $^{13}$CO J = 1 - 0 transitions.
Each field of the GRS in the $^{13}$CO J = 1 - 0 line comprises
spectra on a fully sampled $22''$ grid. The intensities are on a
$T^{*}_{A}$ antenna temperature scale. To convert this to main beam
temperature, $T^{*}_{A}$ is divided by the main beam efficiency of
0.48. The velocity resolution of the data is 0.25 km s$^{-1}$ (0.22
km s$^{-1}$ sampling) \citep{jack2006}. The 15-m JCMT has a FWHM
beam size of $\sim14''$ at $\sim$330 GHz for $^{13}$CO J = 3 - 2
transitions. The correlator was configured with 4096 channels over a
250-MHz bandwidth, which provided a velocity resolution of
$\thicksim$0.055 km s$^{-1}$ per channel. The spectra of $^{13}$CO J
= 1 - 0 and  J = 3 - 2 have been smoothed to a velocity resolution
of 0.50 and 0.44 km s$^{-1}$, respectively.

The $^{13}$CO J = 1 - 0 and the $^{13}$CO J = 3 - 2 data cubes were
processed with CLASS and GREG in the GILDAS software
package\footnote{http://iram.fr/IRAMFR/GILDAS/}.

\section{Results and Discussions}
\label{sect:results}

\subsection{The IR dust distribution of the N68} \label{sec:irstru}

Two Spitzer three-color images of the N68 were shown in
Fig.\ref{Fig:3_colors}. Both figures clearly illustrate the
photo-dissociation region (PDR) visible in 8.0 $\mu$m (in green)
emission, which originates mainly in the polycyclic aromatic
hydrocarbons (PAHs). Due to these large molecules inside the ionized
region are destroyed, the PAH emission delineates the HII region
boundaries, and molecular clouds inside are excited in the PDR by
the radiation leaking from the HII region \citep{petr2010,poma2009}.
The 24 $\mu$m emission (right in Fig.\ref{Fig:3_colors}), appearing
just in the northwest part of the bubble N68, corresponds to hot
dust. It is likely that O- and/or early B-type stars produced the
bubble shell of this HII region, with hot dust located inside the
bubble. The 3.6 $\mu$m emission (in red, left panel in
Fig.\ref{Fig:3_colors}) shows the positions of different stars.

\begin{figure}
\centering
\includegraphics[width=0.49\textwidth, angle=0]{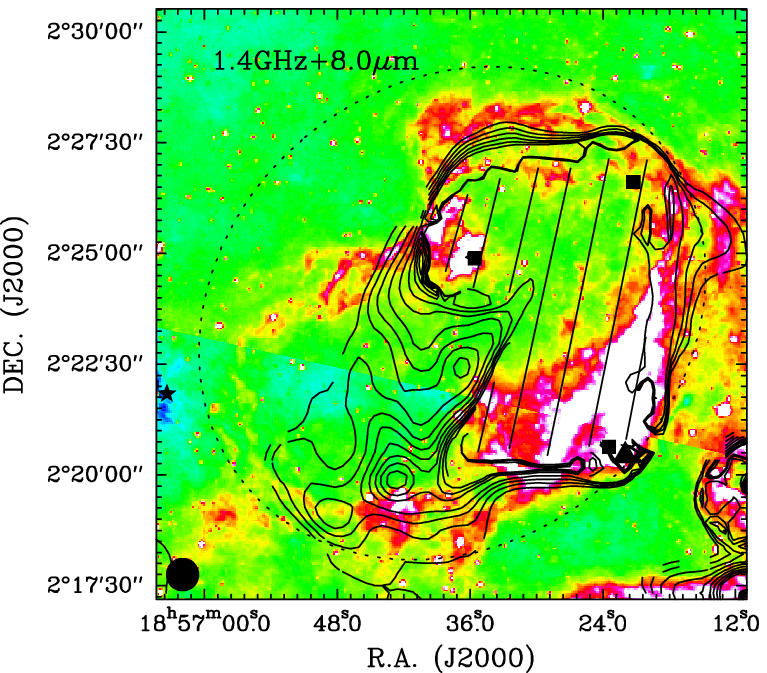}
\includegraphics[width=0.49\textwidth, angle=0]{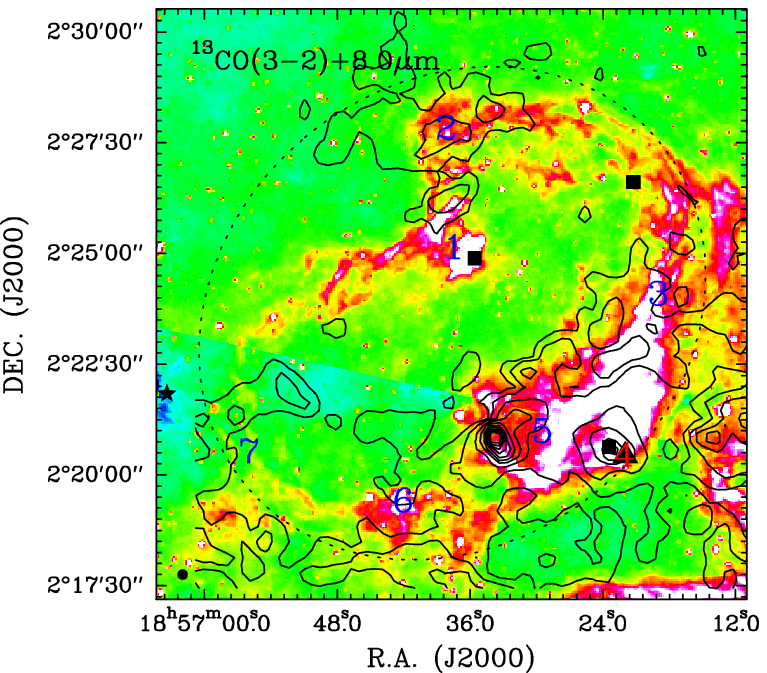}
\caption{Integrated velocity contours of 1.4 GHz continuum (Left
panel) and $^{13}$CO J = 3 - 2 (Right panel), superimposed on the
GLIMPSE 8.0 $\mu$m color images. The contour levels are from 0.16 to
1.45 by 0.16 mJy/beam for 1.4 GHz continuum, while from 6.08, 12.17
to 121.70 by 12.17 K km s$^{-1}$ for $^{13}$CO J = 3 - 2. The
integrated velocity range is from 38.5 to 68.8 km s$^{-1}$ for
$^{13}$CO J = 3 - 2. The symbols ''$\blacktriangle$'',
''$\blacksquare$'', and ''$\bigstar$'' indicate the positions of the
maser (H$_{2}$O and OH), the IRAS point source, and the EGO source,
respectively. The beam size is shown in black filled circle. Note
that the 1.4 GHz continuum are saturated with oblique line in the
image.} \label{Fig:nvss_jcmt}
\end{figure}

In addition, the molecular gas of the bubble shell exhibits several
clumps and filaments along the PDR shell seen from
Fig.\ref{Fig:3_colors}. The distribution and morphology of this
material suggests that a collect-and-collapse process may be
occurring.

\subsection{The continuum emission and clump structure}

The 1.4 GHz continuum NVSS is full of the entire bubble N68 (left
panel in Fig.\ref{Fig:nvss_jcmt}), particularly saturated at the
northwest part. The area, saturated for the NVSS, morphologically
correlates with the 24 $\mu$m emission in Fig.\ref{Fig:3_colors}. We
suggest that exciting stars, which are radiating strong UV
radiation, may exist in this region. Possibly because the
obstruction from molecular cloud, the hot dust develops and
permeates into the southeast part of N68.

The $^{13}$CO J = 3 - 2 contours (Fig.\ref{Fig:nvss_jcmt}) from JCMT
data show several clumps, which may be the birth place of YSOs.
These contours are almost distributed on the 8.0 $\mu$m shell of the
bubble. At the southwest of the N68, there are two strong cores
which will be investigated in the followed Sect.\ref{two_cores}.

\begin{figure}
\centering
\includegraphics[width=0.55\textwidth, angle=0]{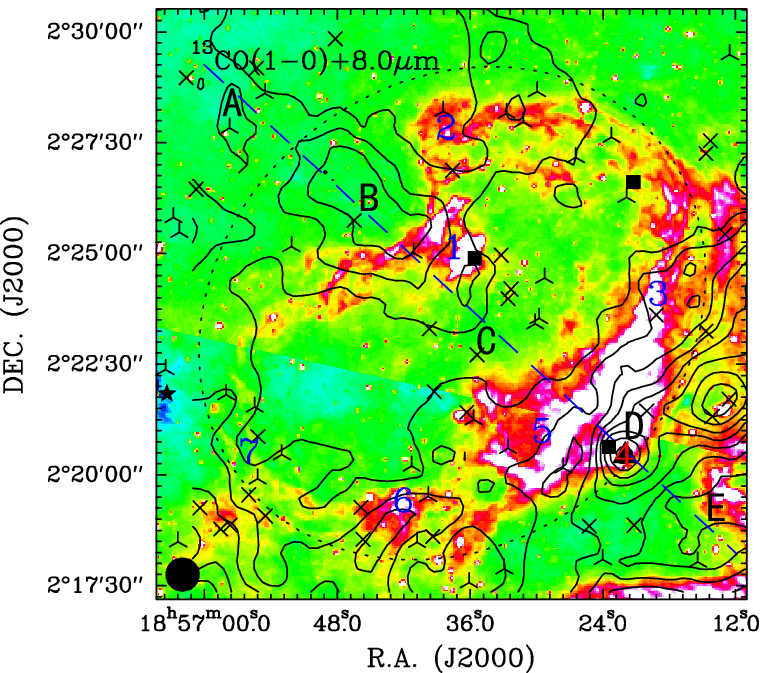}
\includegraphics[width=0.44\textwidth, angle=0]{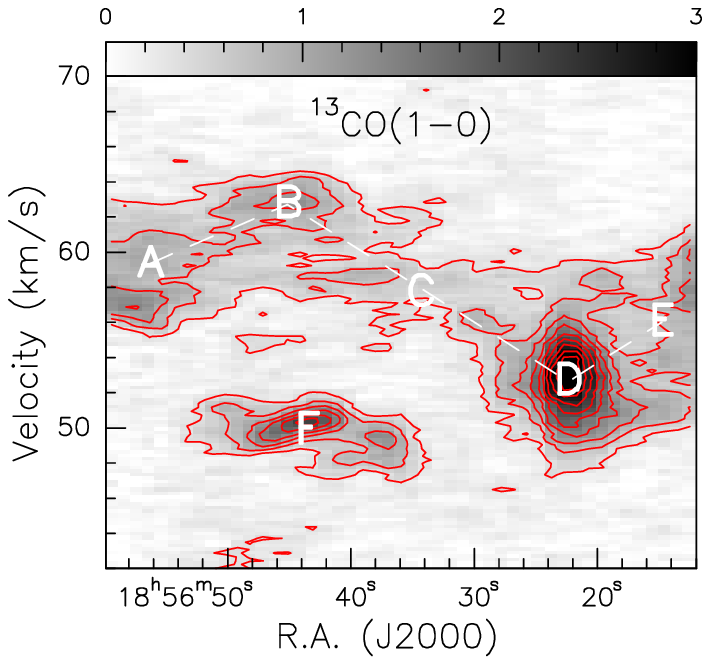}
\caption{Left: integrated velocity contours of the $^{13}$CO J = 1 -
0 emission superimposed on the GLIMPSE 8.0 $\mu$m color image. The
contour levels are from 14.23 to 61.66 by 4.74 K km s$^{-1}$, and
the integrated velocity range is from 38.5 to 68.8 km s$^{-1}$. The
dotted ellipse shows the PAH morphology of bubble. These numbers
give the chosen positions of seven sources. The YSOs candidates of
class I, II are indicated with symbols ''$\times$'', and ''$\Yup$''
respectively. The symbols ''$\blacktriangle$'', ''$\blacksquare$'',
and ''$\bigstar$'' indicate the positions of the maser (H$_{2}$O and
OH), the IRAS point source, and the EGO source, respectively. The
beam size is shown in black filled circle. Right: the
position-velocity diagram of the $^{13}$CO J = 1 - 0 emission along
the dashed line in left panel. The contour levels are from 0.35 to
3.18 by 0.36 K.} \label{Fig:grs}
\end{figure}

The 1.4 GHz continuum flux is about 1.27 Jy within this bubble. The
number of stellar Lyman photon, absorbed by the gas in the HII
region, follows the relation \citep{mezg1974}
\begin{equation}
    \label{eq:nlcy}
    [\frac{N_{LyC}}{s^{-1}}] = 4.761 \times 10^{48}\cdot a(\nu, T_e)^{-1} \cdot [\frac{\nu}{GHz}]^{0.1}\cdot
    [\frac{T_{e}}{K}]^{-0.45}\cdot [\frac{S_{\nu}}{Jy}]\cdot [\frac{D}{kpc}]^2,
\end{equation}
where $a(\nu, T_e)$ is a slowly varying function tabulated by
\citet{mezg1967}; for effective temperature of exciting star $T_{e}
\sim$ 33000 K and at radio wavelengths, $a(\nu, T_e) \sim 1$.
Finally, we obtain Lyman continuum ionizing photons flux log$N_{L}$
$\sim$ 48.79 from nebula. Assuming the exciting stars belong to O9.5
star with log$N_{L}$ = $\sim$ 47.84 \citep{pana1973}, we suggest
there should be about 8.89 exciting stars to ionize the bubble. And
there are presumably many other non-IR excess sources within the
bubble that could also be O- and/or early B-type stars.

\subsection{The distributions of the $^{13}$CO J = 1 - 0 molecular clouds}

\subsubsection{Opacity, column density, and mass}

\begin{figure}
\centering
\includegraphics[width=0.7\textwidth, angle=0]{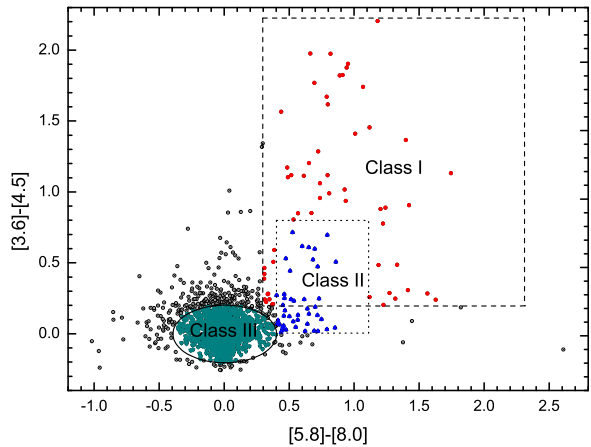}
\caption{The color-color diagram between the [5.8-8.0] and [3.6-4.5]
of the GLIMPSE. Class I, II and III stars of different evolutionary
stage have been indicated in the diagram.} \label{Fig:spitzer}
\end{figure}

We used $^{13}$CO J = 1 - 0 to derive the opacity, column density,
and mass of nebula within the box shown in Fig.\ref{Fig:grs}. The
size of the box is 13.3$'$$\times$13.3$'$. The velocity range is
from $\thicksim$38.5 to $\thicksim$68.8 km s$^{-1}$. The box is
centered on R.A.(J2000) = 18$^{h}$56$^{m}$25$^{s}$.70, DEC.(J2000) =
02$^{\circ}$26$'$01$''$.0.

Here we assume that the nebula is in local thermodynamic equilibrium
(LTE). We used the theory of the radiation transfer and molecular
excitation \citep{winn1979,gard1991}. The $^{13}$CO J = 1 - 0
opacity is gained from the equation:
\begin{equation}
    \label{eq:Tau}
    \tau(^{13}CO)=-ln[1-\frac{T_{MB}(^{13}CO)}{5.3}\{[exp(\frac{5.3}{T_{ex}}-1)]^{-1}-0.16\}^{-1}],
\end{equation}
where the $T_{ex}$ is excitation temperature. The $^{13}$CO J = 1 -
0 column density is obtained from the equation:
\begin{equation}
    \label{eq:N13CO}
    N(^{13}CO)=2.4\times10^{14}\cdot\frac{T_{ex}\int
    \tau(^{13}CO) d v}{1-exp(-5.29/T_{ex})}.
\end{equation}
We assume that the $^{13}$CO J = 1 - 0 is optically thin, the
excitation temperature is $T_{ex}$ = 20 K and the
[H$_{2}$/$^{13}$CO] abundance ratio is $7\times10^{5}$
\citep{frer1982}. The molecular hydrogen column density $N(H_{2})$
was then calculated. The molecular cloud mass is estimated at
$M_{N68} \sim 4.6\times10^{5} M_{\odot}$ including the entire bubble
N68 shell at a distance of $D_{N68} \sim 10.6$ kpc.

\subsubsection{Clump structure}

\begin{figure}
\centering
\includegraphics[width=0.7\textwidth, angle=0]{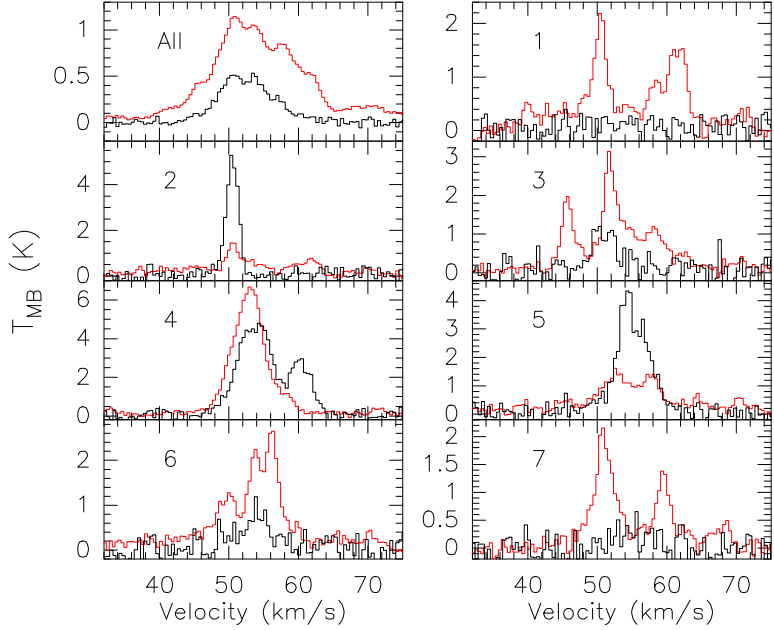}
\caption{$^{13}$CO J = 1 - 0 (red line) and J = 3 - 2 (gray line)
spectra of sources 1 to 7, whose positions are represented in
Fig.\ref{Fig:grs} with number 1 to 7. $All$ indicates that the two
spectra is the spectral average of the bubble N68.}
\label{Fig:core_spectra}
\end{figure}

Fig.\ref{Fig:grs} shows integrated velocity contours of the
$^{13}$CO J = 1 - 0 emission superimposed on the GLIMPSE 8.0 $\mu$m
color image. The integrated velocity range is from 38.5 to 68.8 km
s$^{-1}$. Generally, there should be similar sketch between the
distributions of the millimeter emission and Mid-IR emission
\citep{petr2010,zhan2012}. Comparing the morphologic distribution
(Fig.\ref{Fig:grs}), there are many correlations between the
contours and the 8.0 $\mu$m emission, however, their peak positions
deviate slightly from each other. In addition, the nebula is clumped
and extended outside the bubble. Therefore we have selected seven
sources (1 $\sim$ 7; indicated in Fig.\ref{Fig:grs}) to investigate
their spectral information in followed Sect.\ref{sect:spectra}.

Along the dash-straight line in Fig.\ref{Fig:grs}, we show the
position-velocity diagram. Fig.\ref{Fig:grs} shows that the
components of sources A, B, C, D and E are correlative with the N68,
while the component of source F is not. The components of sources A,
B, C, D and E are made up of a morphology of letter $''N''$ in the
position-velocity diagram (right panel in Fig.\ref{Fig:grs}). Source
C inside bubble N68 has only one velocity component. And from source
B (in the shell of the bubble) to source D (in the shell of the
bubble), the position-velocity diagram dose not show the letter
$''U''$ shape. This is not followed by the model of an expanding
bubble inside a turbulent medium \citep{arce2011}. However,
considering the eccentricity of the bubble and positions of sources
A, B, C, D and E, we also suggest that the bubble N68 may be
expanding outside. The systematic velocity of the N68 is $\sim$58 km
s$^{-1}$ (the source C inside the bubble), and the velocities are
$\sim$59 and $\sim$57 km s$^{-1}$ for sources A and B outside the
bubble, respectively. Also, the velocities are $\sim$63 and $\sim$53
km s$^{-1}$ for sources B and D in the shell of the bubble. So the
bubble N68 may be expanding with a velocity of $\sim$5 km s$^{-1}$
along the line of sight relative to source C.

\begin{figure}
\centering
\includegraphics[width=0.49\textwidth, angle=0]{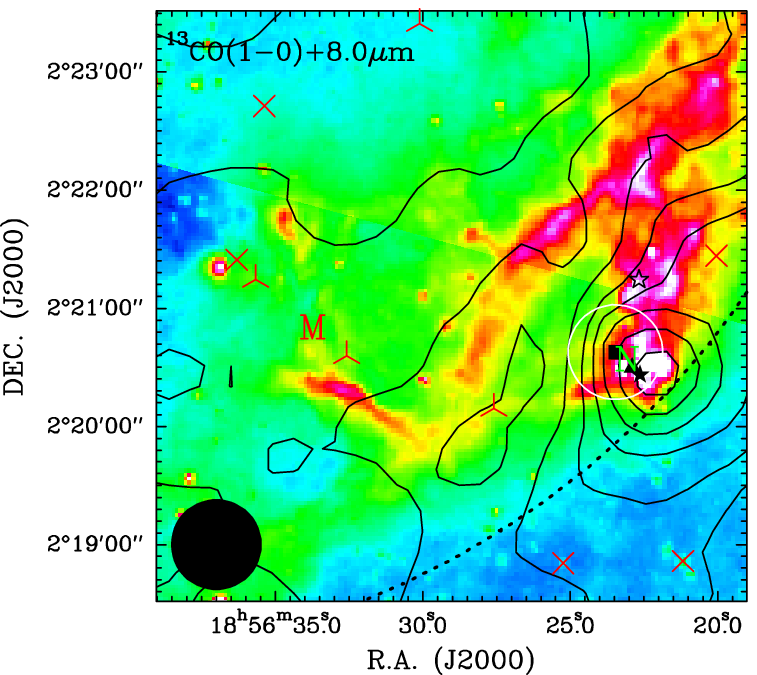}
\includegraphics[width=0.49\textwidth, angle=0]{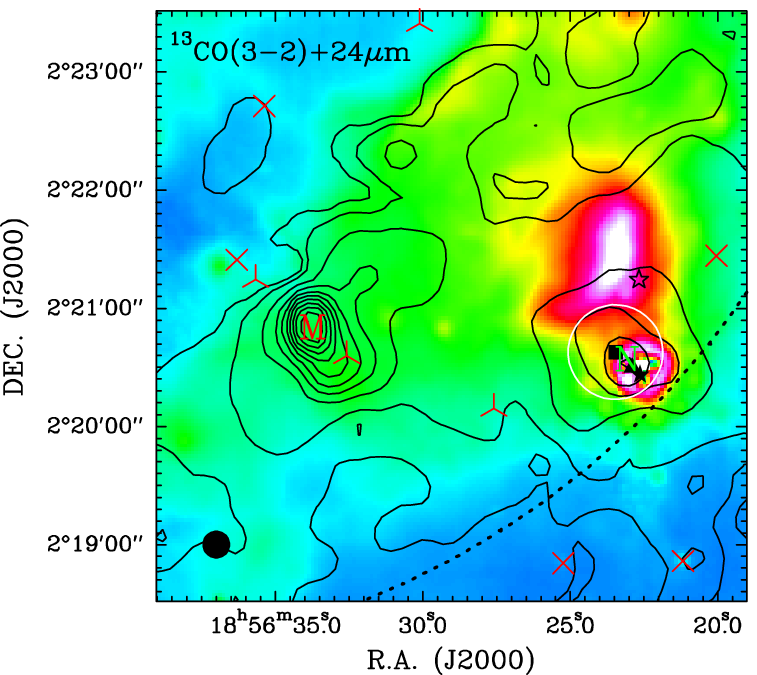}
\caption{Integrated velocity contours of the $^{13}$CO J = 1 - 0 and
J = 3 - 2 emission superimposed on the GLIMPSE 8.0 and 24 $\mu$m
color images, respectively. The contour levels are from 14.23 to
61.66 by 4.74 K km s$^{-1}$ for $^{13}$CO J = 1 - 0, while from
6.65, 13.30 to 119.68 by 13.29 K km s$^{-1}$ for $^{13}$CO J = 3 -
2., and the integrated velocity range is from 38.5 to 68.8 km
s$^{-1}$. The dotted ellipse shows the PAH morphology of bubble.
These numbers give the chosen positions of three sources. The YSOs
candidates of class I, II are indicated with symbols ''$\times$''
and ''$\Yup$''. The symbols ''$\blacktriangle$'',
''$\blacksquare$'', unfilled star, and ''$\bigstar$'' indicate the
positions of the maser (H$_{2}$O and OH), the IRAS point source, the
compact HII region, and the UC HII region, respectively. The white
circle indicates the position error of the IRAS point source.}
\label{Fig:grs_jcmt_small}
\end{figure}

\subsubsection{The young stars distribution}

The color-color diagram (Fig.\ref{Fig:spitzer}) shows the
distribution of class I, II and III stars. Here we only consider
these sources with detection in four Spitzer-IRAC bands
\citep{hora2008}. In addition, to look for a relation between the
young stars and the bubble formation, we draw the class I and II
stars on the 8.0 $\mu$m emission image of the N68
(Fig.\ref{Fig:grs}). We found that class I and II stars almost are
distributed on the shell of the bubble. The distribution of class I
and II stars is also correlative with $^{13}$CO cloud. We also
report the distributions of the IRAS, the extended green object
(EGO), the maser and the UC HII region. These star formation tracers
are located near and on the shell of nebula. It is possible that the
shell of the bubble is the birth place of young stars, which may be
the result of interacting effect between the HII region and
molecular cloud, and be the production of a collect-and-collapse
process.

\begin{figure}
\centering
\includegraphics[width=0.6\textwidth, angle=0]{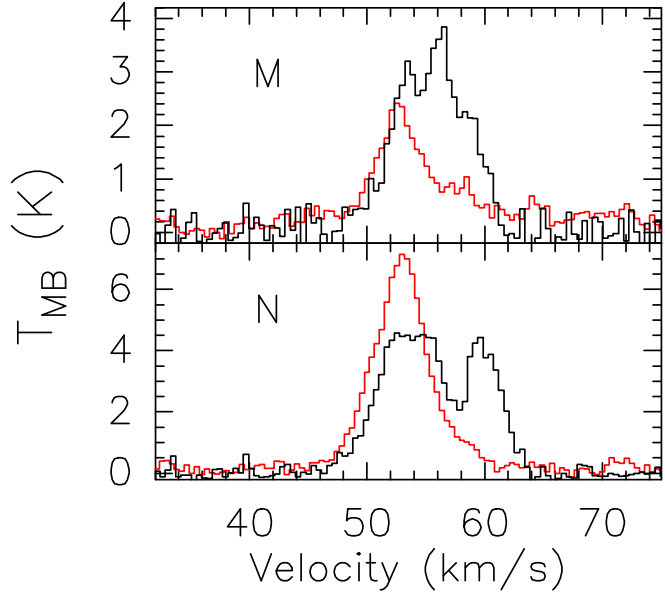}
\caption{$^{13}$CO J = 1 - 0 (red line) and J = 3 - 2 (gray line)
spectra of compact cores M (R.A.(J2000) = 18$^h$56$^m$33$^s$.73,
DEC.(J2000) = 2$^{\circ}$20${'}$49${''}$.81) and N (R.A.(J2000) =
18$^h$56$^m$23$^s$.07, DEC.(J2000) = 2$^{\circ}$20${'}$33${''}$.44),
whose positions are also represented in
Fig.\ref{Fig:grs_jcmt_small}.} \label{Fig:spectra_small}
\end{figure}

\subsubsection{The spectra of the $^{13}$CO J = 1 - 0 and J = 3 -
2}\label{sect:spectra}

Fig.\ref{Fig:core_spectra} shows the $^{13}$CO J = 1 - 0 and J = 3 -
2 spectra of sources 1, 2, 3, 4, 5, 6, and 7 on the shell of the
nebula. The average of all spectra of the whole bubble is exhibited
(indicated with $All$) in panel of Fig.\ref{Fig:core_spectra}. This
shows that the spectra of the $^{13}$CO J = 1 - 0 has a wider width
and stronger intensity than those of the J = 3 - 2. At the sources 2
and 5, however, the spectra of the $^{13}$CO J = 3 - 2 has a
stronger intensity than that of the J = 1 - 0. The $^{13}$CO J = 1 -
0 spectrum of source 4 does not have the $\sim$61 km s$^{-1}$
component relative to the J = 3 - 2. In addition, from the contours,
the $^{13}$CO J = 3 - 2 distribution (Fig.\ref{Fig:grs}) is clumpier
than that of the J = 1 - 0 (Fig.\ref{Fig:nvss_jcmt}). Therefore, the
difference between the $^{13}$CO J = 1 - 0 and J = 3 - 2 is very
much, and needs further investigation.

The intensity ratio of the $^{13}$CO J = 3 - 2 to J = 1 - 0 is
$R^{13}_{3-2/1-0}$ = 0.46, 3.53, 0.38, 0.70, 2.69, and 0.46 for
sources $All$, 2, 3, 4, 5, and 6, respectively. Our results are
consistent with that of \citet{mina2011}, whose $R^{13}_{3-2/1-0}$
ranges from 3.6$\pm$0.9 to 0.24$\pm$0.07. This is suggestive that
$^{13}$CO J = 1 - 0 and J = 3 - 2 transitions have different
critical densities, so the $R^{13}_{3-2/1-0}$ traces the excitation
of molecular gas.

\subsection{Two compact cores in the shell of the N68}\label{two_cores}

There are two compact cores M (R.A.(J2000) = 18$^h$56$^m$33$^s$.73,
DEC.(J2000) = 2$^{\circ}$20${'}$49${''}$.81) and N (R.A.(J2000) =
18$^h$56$^m$23$^s$.07, DEC.(J2000) = 2$^{\circ}$20${'}$33${''}$.44),
located at the southwest shell of N68. Fig.\ref{Fig:grs_jcmt_small}
shows the integration contours of the $^{13}$CO J = 1 - 0 and J = 3
- 2 emission. The cores M and N are located at the peaks of
integration contours of the $^{13}$CO J = 3 - 2 emission. The peak
of the $^{13}$CO J = 1 - 0 contours is located at core N, but it
does not appear for core M. In addition, the difference between the
8.0 and 24 $\mu$m emission is obvious. The present of the 24 $\mu$m
emission is suggestive that the hot dust has plunged into the bubble
shell.

On the other hand, there are many star formation tracers located
near core N. We marked the positions of a H$_{2}$O maser, a OH
maser, a IRAS point source, a compact HII region and a UC HII region
in Fig.\ref{Fig:grs_jcmt_small}. The H$_{2}$O maser, the OH maser,
the IRAS point source and the UC HII region are located at the core
N. There are many velocity components for the H$_{2}$O maser and the
OH maser. We just list three of their strongest components. The
fluxes are 20.15 Jy at 48.90 km s$^{-1}$, 7.25 Jy at 49.45 km
s$^{-1}$ and 2.80 Jy 48.35 km s$^{-1}$ for H$_{2}$O maser emission,
while 9.90 Jy at 50.00 km s$^{-1}$, 4.40 Jy 46.05 km s$^{-1}$ and
3.70 Jy at 47.36 km s$^{-1}$ for the OH maser emission
\citep{fors1999}. These are suggestive that core N is a strongly
active massive star forming region.

We also exhibit the spectra of cores M and N in
Fig.\ref{Fig:spectra_small}. The difference of spectral profiles
between $^{13}$CO J = 1 - 0 and J = 3 - 2 emission is obvious. the
$^{13}$CO J = 3 - 2 has double profiles for cores M and N, but the
$^{13}$CO J = 1 - 0 lacks of the blue profile. For core N, the
$^{13}$CO J = 1 - 0 has a main beam temperature of $\sim$7 K at
$\sim$53 km s$^{-1}$, yet the $^{13}$CO J = 3 - 2 shows a flat
spectrum of absorption. This discrepancy is needed to further
investigate.

\section{Conclusions}
\label{sect:summary}

We have investigated the environment of the IR dust bubble N68 with
the $^{13}$CO J = 1 - 0 and J = 3 - 2 emission lines, the 1.4 GHz
continuum, and the IR bands emission. The main results can be
summarized as follows.

The morphology of the molecular line emission ($^{13}$CO J = 1 - 0
and J = 3 - 2) and the associated velocity signatures are consistent
with the shell structure seen from the 8.0 $\mu$m images. The
expanding shells in N68 are suggestive of triggered massive star
formation. There is evidence for dense clumps coincident with the
shells. The expanding shells have an expansion speed of $\sim$5 km
s$^{-1}$. However, it is not clear whether HII region is the driving
mechanism of the shell expansion.

We have presented a study about compact cores M and N located at the
shell of nebula. Compact core N is associative with the H$_{2}$O
maser, the OH maser, the IRAS point source, and the UC HII region.
The spectral velocity components of the H$_{2}$O maser and the OH
maser indicate that the star formation region is strongly activate.
However, the deep-going properties of the compact cores M and N are
needed to further investigate.

We also used the GLIMPSE, and the MIPSGAL survey data to analyze the
YSOs and the warm dust distribution around the bubble N68. We
identified class I and II YSOs using the [5.8]-[8.0] versus
[3.6]-[4.5] relation, and correlated their distribution relative to
the PDR, which we assume to be associated with and surrounding an
HII region. We find that the N68 appears to have a significant
number of YSOs associated with their PDRs, implying that triggered
star formation mechanisms acting on the boundary of the expanding
HII region.

\section{Acknowledgements}

This work was supported by the Young Researcher Grant of National
Astronomical Observatories, Chinese Academy of Sciences
No.O835032002.

\bibliographystyle{raa}
\bibliography{references}

\end{document}